\documentclass[conference, compsoc]{IEEEtran}

\newif\ifExtended
\Extendedfalse

\usepackage{booktabs} 
\usepackage{tikz}
\usetikzlibrary{shadows}
\usepackage{pgfplots}
\pgfplotsset{compat=1.10}

\usepackage{colortbl}
\definecolor{Gray}{gray}{0.925}

\usepackage{algorithm}
\usepackage{algorithmicx}
\usepackage{algpseudocode}
\usepackage[inline]{enumitem}
\usepackage{comment}
\usepackage{soul}
\usepackage{xspace}

\usepackage{amssymb}
\usepackage{amsmath}

\newcommand{\vect}[1]{\ensuremath{\boldsymbol{#1}}}
\newcommand{\va}{\vect{a}}
\newcommand{\vo}{\vect{o}}

\newcommand{\Natural}[0]{\ensuremath{\mathbb{N}}}
\newcommand{\Real}[0]{\ensuremath{\mathbb{R}}}

\usepackage{amsthm}

\theoremstyle{definition}
\newtheorem{definition}{Definition}

\usepackage{listings, xcolor}
\definecolor{verylightgray}{rgb}{.97,.97,.97}
\lstdefinelanguage{Solidity}{
	keywords=[1]{anonymous, assembly, assert, balance, break, call, callcode, case, catch, class, constant, continue, contract, debugger, default, delegatecall, delete, do, else, event, export, external, false, finally, for, function, gas, if, implements, import, in, indexed, instanceof, interface, internal, is, length, library, log0, log1, log2, log3, log4, memory, modifier, new, payable, pragma, private, protected, public, pure, push, require, return, returns, revert, selfdestruct, send, storage, struct, suicide, super, switch, then, this, throw, transfer, true, try, typeof, using, value, view, while, with, addmod, ecrecover, keccak256, mulmod, ripemd160, sha256, sha3}, 
	keywordstyle=[1]\color{blue}\bfseries,
	keywords=[2]{address,..., bool, byte, bytes, bytes1, bytes2, bytes3, bytes4, bytes5, bytes6, bytes7, bytes8, bytes9, bytes10, bytes11, bytes12, bytes13, bytes14, bytes15, bytes16, bytes17, bytes18, bytes19, bytes20, bytes21, bytes22, bytes23, bytes24, bytes25, bytes26, bytes27, bytes28, bytes29, bytes30, bytes31, bytes32, enum, int, int8, int16, int24, int32, int40, int48, int56, int64, int72, int80, int88, int96, int104, int112, int120, int128, int136, int144, int152, int160, int168, int176, int184, int192, int200, int208, int216, int224, int232, int240, int248, int256, mapping, string, uint, uint8, uint16, uint24, uint32, uint40, uint48, uint56, uint64, uint72, uint80, uint88, uint96, uint104, uint112, uint120, uint128, uint136, uint144, uint152, uint160, uint168, uint176, uint184, uint192, uint200, uint208, uint216, uint224, uint232, uint240, uint248, uint256, var, void, ether, finney, szabo, wei, days, hours, minutes, seconds, weeks, years},	
	keywordstyle=[2]\color{teal}\bfseries,
	keywords=[3]{block, blockhash, coinbase, difficulty, gaslimit, number, timestamp, msg, data, gas, sender, sig, value, now, tx, gasprice, origin},	
	keywordstyle=[3]\color{violet}\bfseries,
	identifierstyle=\color{black},
	sensitive=false,
	comment=[l]{//},
	morecomment=[s]{/*}{*/},
	commentstyle=\color{gray}\ttfamily,
	stringstyle=\color{red}\ttfamily,
	morestring=[b]',
	morestring=[b]"
}
\usepackage{caption}
\usepackage{subcaption}
\usepackage[textsize=tiny, textwidth=1.5cm,disable]{todonotes}

\newcommand{\ad}[1]{\todo[color=yellow!50, linecolor=black!50]{\textbf{Abhishek}: #1}}
\newcommand{\Aron}[1]{\todo[color=blue!25, linecolor=black!50]{\textbf{Aron}: #1}}

\newcommand{\Scott}[1]{\todo[color=orange!30, linecolor=black!50]{\textbf{Scott}: #1}}

\usepackage{url}
\graphicspath{{./figs/}}
\newcommand{\subparagraph}{}

\usepackage{titlesec}

\titlespacing\section{0pt}{8pt plus 4pt minus 2pt}{4pt plus 2pt minus 2pt}
\titlespacing\subsection{0pt}{8pt plus 4pt minus 2pt}{4pt plus 2pt minus 2pt}
\titlespacing\subsubsection{0pt}{8pt plus 4pt minus 2pt}{4pt plus 2pt minus 2pt}

\newcommand{\Platform}{\textit{SolidWorx}\xspace}

\begin{document}

\setlength{\marginparwidth}{1.4cm}

\thispagestyle{plain}
\pagestyle{plain}

\title{\Platform: A Resilient and Trustworthy Transactive Platform for\\Smart and Connected Communities}
\author{
\IEEEauthorblockN{Scott Eisele}
\IEEEauthorblockA{Vanderbilt University \\
Nashville, TN, USA \\ 
scott.r.eisele@vanderbilt.edu
}

\and
\IEEEauthorblockN{Aron Laszka}
\IEEEauthorblockA{University of Houston\\
Houston, TX, USA \\
alaszka@houston.edu}
\and
\IEEEauthorblockN{Anastasia Mavridou}
\IEEEauthorblockA{Vanderbilt University\\
Nashville, TN, USA \\
anastasia.mavridou@vanderbilt.edu}
\and
\IEEEauthorblockN{Abhishek Dubey}
\IEEEauthorblockA{Vanderbilt University \\
Nashville, TN, USA \\ 
abhishek.dubey@vanderbilt.edu
}
}

\maketitle

\begin{abstract}
Internet of Things and data sciences are fueling the development of innovative solutions for various applications in Smart and Connected Communities (SCC). These applications provide participants with the capability to exchange not only data but also resources, which raises the concerns of integrity, trust, and above all the need for fair and optimal solutions to the problem of resource allocation. This exchange of information and resources  leads to a problem where the stakeholders of the system may have limited trust in each other. Thus, collaboratively reaching consensus on when, how, and who should access certain resources becomes problematic. This paper presents \Platform, a blockchain-based platform that provides key mechanisms required for  arbitrating resource consumption across different SCC applications in a domain-agnostic manner. For example, it introduces and implements a hybrid-solver pattern, where  complex optimization computation is handled off-blockchain while solution validation is performed by a smart contract. To ensure correctness, the smart contract of \Platform is generated and verified.

\end{abstract}
\setlength{\belowcaptionskip}{-10pt}
\section{Introduction}
\label{sec:intro}

%

%
Smart and connected communities (SCC) as a research area lies at the intersection of social science, machine learning, cyber-physical systems, civil infrastructures, and data sciences. This research area is  enabled by the rapid and transformational changes driven by  innovations in smart sensors, such as cameras and air quality monitors, which are now  embedded in almost every physical device and system that we use, ranging from watches and smartphones to automobiles, homes, roads, and workplaces. 
The effects of these innovations can be seen in a number of diverse domains, including transportation, energy, emergency response, and  health care, to name a few.

At its core, a smart and connected community is a
multi-agent system where agents may enter or leave the system for different reasons. Agents may act on behalf of service owners, managing access to services and ensuring that contracts are fulfilled. Agents can also act on behalf of service consumers, locating services, making contracts, as well as receiving and presenting results.
For example, agents may coordinate carpooling services. Another example of such coordination exists in transactive energy systems~\cite{Gridwise}, where homeowners in a community exchange excess energy. Consequently, these  agents are required to engage in interactions, negotiate with each other, enter agreements, and make proactive run-time decisions---individually and collectively---while responding to changing circumstances. 

This exchange of information and resources  leads to a problem where the stakeholders of the system may have limited trust in each other. Thus, collaboratively reaching consensus on when, how, and who should access certain resources becomes problematic. However, instead of solving these problems in a domain specific manner, we present \Platform  and show how this platform can provide key design patterns to implement mechanisms for arbitrating resource consumption across different SCC applications. 

Blockchains may form a key component of SCC platforms because they enable participants to reach a consensus on the value of any state variable in the system, without relying on a trusted third party or trusting each other. Distributed consensus not only solves the trust issue, but also provides fault-tolerance since consensus is always reached on the correct state as long as the number of faulty nodes is below a threshold. Further, blockchains can also enable performing computation in a distributed and trustworthy manner in the form of smart contracts. However, while the distributed integrity of a blockchain ledger presents unique opportunities, it also introduces new assurance challenges that must be addressed before protocols and implementations can live up to their potential. For instance, Ethereum smart contracts deployed in practice are riddled with bugs and security vulnerabilities.  Thus, we use a correct-by-construction design toolchain, called FSolidM \cite{mavridou2018designing}, to design and implement  the smart-contract code of \Platform. 

The outline of this paper is as follows. We formulate a resource-allocation problem for SCC in Section~\ref{sec:problem},  describing two concrete  applications of the platform in Section~\ref{sec:ExaApp} and presenting extensions to the basic problem formulation in Section~\ref{sec:ProForExt}. We describe our solution architecture in Section~\ref{sec:solution}, which consists of off-blockchain solvers (Section~\ref{sec:solver}) and a smart contract (Section~\ref{sec:smartcontract}), providing a brief analysis in Section~\ref{sec:analysis}. In Section~\ref{sec:results}, we evaluate \Platform using two case studies, a carpooling assignment (Section~\ref{sec:carpool})
 and an energy trading system  (Section~\ref{sec:energy}). Finally, we discuss the architecture of \Platform in the context of related research in Section~\ref{sec:related},  and we provide concluding remarks in Section \ref{sec:conclusion}. 




\section{Problem Formulation}
\label{sec:problem}

We first introduce a base formulation of an abstract resource allocation problem (Section~\ref{sec:ResAllPro}), which captures the core functionality of a transactive platform for SCC.
Then, we describe two  examples of applying this formulation to solving practical problems in SCC (Section~\ref{sec:ExaApp}).
We conclude the section by introducing various extensions to the base problem formulation, in the form of alternative objectives and additional constraints (Section~\ref{sec:ProForExt}).
A list of the key symbols used in the resource allocation problem can be found in Table~\ref{tab:symbols}.

\begin{table}[t]
    \centering
    \caption{List of Symbols}
    \label{tab:symbols}
    \renewcommand{\arraystretch}{1.2}
    \begin{tabular}{|c|p{6.86cm}|}
        \hline
        Symbol & Description \\
        \hline
        \rowcolor{Gray} $P$ & set of resource providers \\
        $C$ & set of resource consumers \\
        \rowcolor{Gray} $T$ & set of resource types \\
        $O\!P$ & set of providing offers \\
        \rowcolor{Gray} $OC$ & set of consumption offers \\
        $o_P$ & resource provider who posted offer $\vo \in O\!P$ \\
        \rowcolor{Gray} $o_C$ & resource consumer who posted offer $\vo \in OC$ \\
        $o_Q(t)$ & amount of resources of type $t \in T$ provided or requested by offer $\vo$ \\
        \rowcolor{Gray} $o_V(t)$ & unit reservation price of offer $\vo$ for resource type $t \in T$ \\
        $a_{O\!P}$ & providing offer from which assignment $\va$ allocates resources \\
        \rowcolor{Gray} $a_{OC}$ & consuming offer to which assignment $\va$ allocates resources \\
        $a_Q$ & amount of resources allocated by assignment $\va$ \\
        \rowcolor{Gray} $a_T$ & type of resources allocated by assignment $\va$ \\
        $a_V$ & unit price for the resources allocated by assignment $\va$ \\
        \hline
    \end{tabular}
\end{table}

\subsection{Resource Allocation Problem}
\label{sec:ResAllPro}

In essence, the objective of the transactive platform is to allocate resources from users who provide resources to users who consume them.
The sets of \emph{resource providers} and \emph{resource consumers} are denoted by $P$ and $C$, respectively.
Note that a user may act both as a resource provider and as a resource consumer at the same time, in which case the user is a member of both $P$ and~$C$. 
Resources that are provided or consumed belong to a set of \emph{resource types}, which are denoted by $T$.
A resource type is an abstract concept, which captures not only the inherent characteristics of a resource, but all aspects related to providing or consuming resources.
For example, a resource type could correspond to energy production and consumption in a specific time interval, or it could correspond to a ride between certain location at a certain time.

Each provider $p \in P$ may post a set of \emph{providing offers}. 
Each providing offer $\vo$ 
is a tuple $\vo = \langle o_P, o_Q, o_V \rangle$, where $o_P \in P$ is the provider who posted the offer, $o_Q \in T \mapsto \Natural$ is the amount of resources offered from each type (i.e., $o_Q(t)$ is the amount of resources offered from type $t \in T$), and $o_V \in T \mapsto \Natural$ is the unit reservation price asked for each resource type (i.e., $o_V(t)$ is the value asked for providing a unit resource of type $t \in T$).
Each offer $\vo = \langle o_P, o_Q, o_V \rangle$ defines a set of alternatives: provider $o_p$ offers to provide either $o_Q(t_1)$ resources of type $t_1 \in T$ or $o_Q(t_2)$ resources of type $t_2 \in T$, but not at the same time.
However, convex linear combinations, such as providing $\lfloor \alpha \cdot o_Q(t_1)  \rfloor$ resources of type $t_1 \in T$ and $\lfloor (1 - \alpha) \cdot o_Q(t_2) \rfloor$ resources of type $t_2 \in T$ at the same time (where $\alpha \in [0, 1]$), are allowed.
For example, an offer $\vo$ providing $o_Q(t_1)$ units of energy in time interval~$t_1$ or $o_Q(t_2)$ units of energy in time interval $t_2$ may provide $\lfloor 0.5 \cdot o_Q(t_1) \rfloor$ energy in time interval $t_1$ and $\lfloor 0.5 \cdot o_Q(t_2) \rfloor$ energy in time interval $t_2$.
The set of all offers posted by all the providers is denoted by $O\!P$. 

Each consumer $c \in C$ posts a set of \emph{consumption offers}. 
Each consumption offer $\vo$ 
is a tuple $\vo = \langle o_C, o_Q, o_V \rangle$, where $o_C \in C$ is the consumer who posted the offer, $o_Q \in T \mapsto \Natural$ is the amount of resources requested from each type (i.e., $o_Q(t)$ is the amount of resources requested from type $t \in T$), and $o_V \in T \mapsto \Natural$ is the unit reservation price offered for each resource type (i.e., $o_V(t)$ is the value offered for a unit resource of type $t \in T$).
Similar to providing offers, consumption offers also define a set of alternatives.
The set of all offers posted by all the consumers is denoted by $OC$. 

A \emph{resource allocation} $A$ is a set of resource assignments.
Each resource assignment $\va \in A$ is a tuple $\va = \langle a_{O\!P}, a_{OC}, a_Q, a_T, a_V \rangle$, where $a_{O\!P} \in O\!P$ is a providing offer posted by a provider, $a_{OC} \in OC$ is a consumption offer posted by a consumer, $a_Q \in \Natural$ and $a_T \in T$ are the amount and type of resources allocated from offer $a_{O\!P}$ to $a_{OC}$, and $a_V \in \Natural$ is the unit price for the assignment.

A resource allocation $A$ is \emph{feasible} if
\begin{align}
\forall \vo \in O\!P: ~ & \sum_{t \in T} ~ \sum_{\substack{\va \in A:\\a_{O\!P} = \vo \,\wedge\, a_T = t}} \frac{a_Q}{o_Q(t)} \leq 1 \label{eq:feasible1} \\
\forall \vo \in OC: ~ & \sum_{t \in T} ~ \sum_{\substack{\va \in A:\\a_{OC} = \vo \,\wedge\, a_T = t}} \frac{a_Q}{o_Q(t)} \leq 1 \label{eq:feasible2} \\
\forall \va \in A: ~ & {\left(a_{O\!P}\right)}_V(a_T) \leq a_V \\
\forall \va \in A: ~ & {\left(a_{OC}\right)}_V(a_T) \geq a_V \label{eq:feasible4} .
\end{align}

In other words, a resource allocation is feasible if the resources assigned from each providing offer (or consuming offer) is a convex linear combination of the offered (or requested) resources, and if the value in each assignment is higher than (or lower than) the reservation price of the providing offer (or consuming offer).

The objective of the base formulation of the \emph{resource allocation problem} is to maximize the amount of resources assigned from providers to consumers.
We define the base formulation of the problem as follows.

\begin{definition}[Resource Allocation Problem]
Given sets of providing and consumption offers $O\!P$ and $OC$, find a feasible resource allocation $A$ that attains the maximum
\begin{equation}
\max_{A:\, A \textnormal{ is feasible}} \, 
\sum_{\va \in A} a_Q . \label{eq:objective}
\end{equation}
\end{definition}

\subsection{Example Applications}
\label{sec:ExaApp}

To illustrate how the Resource Allocation Problem (RAP) may be applied in smart and connected communities, we now describe two example problems that can be expressed using RAP.

\subsubsection{Energy Futures Market}
\label{sec:energyFuturesMarket}

\newcommand{\etime}[0]{\ensuremath{t}}

We consider a residential energy-futures market in a transactive microgrid.
In this application, resource consumers model residential energy consumers (i.e., households), while resource providers model the subset of consumers who have energy providing capabilities (e.g., solar panels, batteries).
We divide time into fixed-length intervals (e.g., 15 minutes),
 and let each resource type correspond to providing or consuming a unit amount of power (e.g., 1 W) in a particular time interval.

Based on their predicted energy supply and demand, residential consumers (or smart homes acting on their behalf) post offers to provide or consume energy in future time intervals.
For instance, a provider may predict that it will be able to generate a certain amount of power $\pi$ using its solar panel during time intervals $\etime_1, \etime_2, \ldots, \etime_N \in T$.
Then, it will submit a \emph{set of $N$ offers}: for each time interval~$\etime \in \{\etime_1, \ldots, \etime_N\}$ in which energy may be produced, it posts an offer specifying 
\begin{equation}
    o_Q(t) = \begin{cases}
    \pi & \textnormal{ if } t = \etime \\
    0 & \textnormal{ otherwise.}
    \end{cases}
\end{equation}
Alternatively, the provider may have a fully charged battery, which could be discharged in any of the next $N$ intervals $\etime_1, \etime_2 \ldots, \etime_N$.
Let $\pi$ denote the amount of power that could be provided if the battery was fully discharged in a single time interval.
Then, the provider will submit a \emph{single offer} specifying
\begin{equation}
    o_Q(t) = \begin{cases}
    \pi & \textnormal{ if } t \in \left\{ \etime_1, \etime_2, \ldots, \etime_N \right\} \\
    0 & \textnormal{ otherwise.}
    \end{cases}
\end{equation}

The reservation prices of the offers should consider the energy prices of the utility company (i.e., the alternative to local trading) and 
the cost of providing energy (e.g., cost of battery depreciation due to charging and discharging).


\subsubsection{Carpooling Assignment}
\label{sec:carpoolingprob}
We consider the problem of assigning carpooling riders to drivers with empty seats in their cars.
In this application, resource consumers model riders, while resource providers model drivers.
We again divide time into fixed-length intervals, and we divide the space of pick-up locations into a set of areas (e.g., city blocks).
Then, we let a resource type correspond to a ride from a particular area in a particular time interval to a particular area. 
A unit of a resource is a single seat for a ride.

A provider (i.e., driver) who has $\pi$ empty seats in its car will post a providing offer. 
Let $\Pi \subseteq T$ denote the set of   combinations of pick-up and drop-off areas and pick-up times that are feasible for the provider.
Then, the provider's offer specifies
\begin{equation}
    o_Q(t) = \begin{cases}
    \pi & \textnormal{ if } t \in \Pi \\
    0 & \textnormal{ otherwise.}
    \end{cases}
\end{equation}
Similarly, a consumer (i.e., rider) who needs 1 seat will post a consuming offer, specifying
\begin{equation}
    o_Q(t) = \begin{cases}
    1 & \textnormal{ if } t \in \Pi \\
    0 & \textnormal{ otherwise,}
    \end{cases}
\end{equation}
where $\Pi$ is the set of combinations (i.e., pick-up and drop-off areas and pick-up times) that are feasible for the rider.

\subsection{Problem Formulation Extensions}
\label{sec:ProForExt}

\Aron{Note to self: review this subsection!}
The Resource Allocation Problem that we introduced in Section~\ref{sec:ResAllPro} can capture a wide range of real-world problems.
However, some problems may not be easily expressed using the constraints (Equations \eqref{eq:feasible1} to \eqref{eq:feasible4}) and the objective (Equation \eqref{eq:objective}) of the base problem formulation.
For this reason, here we introduce a set of alternative objective formulations and additional constraints for resource allocation. 

\subsubsection{Objectives}
\label{sec:extObj}

We first introduce alternative objective formulations, which quantify the utility of a resource allocation based on alternative goals.

\textbf{Resource Type Preferences:}
Equation \eqref{eq:objective} assumes that exchanging a unit of any resource type is equally beneficial.
In some practical scenarios, exchanging certain resource types may be more beneficial than exchanging others.
For each resource type~$t \in T$, let $\beta_t$ denote the utility derived from exchanging a unit of resources of type $t$.
Then, the utility of a resource allocation~$A$ can be expressed as
\begin{equation}
    \sum_{\va \in A} \beta_{\left(a_T\right)} \cdot a_Q .
\end{equation}

\textbf{Provider and Consumer Benefit:}
The reservation price~$o_V(t)$ of a providing offer $\vo$ means that provider $o_P$ is indifferent to (i.e., derives zero benefit from) exchanging resources of type $t$ at unit price $o_V(t)$.
Hence, the unit benefit derived by the provider from exchanging at a higher price $a_V \geq o_V(t)$ is equal to $a_V - o_V(t)$.
Similarly, the unit benefit derived by a consumer, who posted an offer $\vo$, from exchanging resources of type $t$ at price $a_V$ is equal to $o_V(t) - a_V$.
Therefore, the total benefit created by an assignment $\va$ for provider $a_{O\!P}$ and consumer $a_{OC}$ is
\begin{align}
    & a_Q \cdot \left[ a_V - \left(a_{O\!P}\right)_V(a_T) \right] + a_Q \cdot \left[ \left(a_{OC}\right)_V(a_T) - a_V \right] \nonumber \\
    = & a_Q \cdot \left[ \left(a_{OC}\right)_V(a_T) - \left(a_{O\!P}\right)_V(a_T) \right] ,
\end{align}
and the total benefit created by a resource allocation $A$ for all the providers and consumers is
\begin{equation}
    \sum_{\va \in A}  a_Q \cdot \left[ \left(a_{OC}\right)_V(a_T) - \left(a_{O\!P}\right)_V(a_T) \right] .
\end{equation}

\subsubsection{Constraints}
\label{sec:extConstr}

Next, we introduce additional feasibility constraints that may be imposed on the resource allocations.

\textbf{Price Constraints:}
A regulator (e.g., utility company in a transactive energy platform) may impose constraints on the prices at which resources may be exchanged (e.g., based on bulk-market prices).
If the minimum and maximum unit prices for resource type $t \in T$ are $min_t$ and $max_t$, respectively, then we can express price constraints as
\begin{equation}
    \forall \va \in A: ~ min_{\left( a_T \right)} \leq a_V \leq max_{\left( a_T \right)} . 
\end{equation}

\textbf{Pairwise Constraints:}
Due to physical constraints, exchanging resources of certain types between certain pairs of prosumers may be impossible.
If the set of prosumer pairs that may exchange resources of type $t \in T$ is denoted by~$E_t \subseteq P \times C$,   we can express pairwise constraints as
\begin{equation}
    \forall \va \in A: \left( a_{O\!P} , a_{OC} \right) \in E_{\left( a_T \right)} .
\end{equation}

\textbf{Real-Valued Offers and Allocations:}
Finally, we may also relax some of the constraints of the base formulation.
In particular, we may allow real-valued quantities in offers and allocations (i.e., $o_Q: T \mapsto \Real_+$ and $a_Q \in \Real_+$) as well as real-valued prices (i.e., $o_V: T \mapsto \Real_+$ and $a_V \in \Real_+$).


\section{\Platform: A Decentralized Transaction Management Platform}
\label{sec:solution}

Now, we describe the \Platform platform, which (1) allows prosumers\footnote{An actor or an agent that can both provide and consume resources.} to post offers and (2) can find a solution to the resource allocation problem in an efficient and trustworthy manner. SolidWorx follows the actor-based architecture, which was proposed initially in~\cite{agha1985actors}, and which has been accepted as a standard model for building distributed applications. The key aspect of an actor-based system are interfaces with well defined execution models~\cite{basu2011rigorous}. 
The following subsections will describe the transaction management platform in more detail. Here, we provide a brief overview.

\begin{figure}[t]
    \centering
   

\begin{tikzpicture}[x=1.2cm, y=1.2cm, font=\tiny,
  Component/.style={fill=white, draw, align=center, rounded corners=0.1cm, drop shadow={shadow xshift=0.05cm, shadow yshift=-0.05cm, fill=black}},
  Connection/.style={<->, >=stealth, shorten <=0.05cm, shorten >=0.05cm}]

\foreach \pos/\name in {1.6/pros3, 0.8/pros2} { 
  \node [Component] (\name) at (\pos - 4, \pos) {\textbf{Prosumer}\\(\texttt{Python}, \texttt{geth})};
}

\node [Component] (dso) at (-1, 2.6) {\textbf{Directory}\\(\texttt{Python}, \texttt{geth})};

\node [Component] (solver) at (-4, 0) {\textbf{Solver}\\(\texttt{Python}, \texttt{CPLEX}, \texttt{geth})};

\fill [fill=black!15] (90:1.5) -- (200:1.5) -- (340:1.5) -- (90:1.5);

\foreach \pos in {90, 200, 340} {
  \node [Component] at (\pos:1.5) {Blockchain\\ miner (\texttt{geth})};
}

\node [Component, dotted] (contract) at (0, 0) {\textbf{Smart Contract}\\(\texttt{Solidity})};

\draw [Connection, bend left=62] (solver) to node [midway, left, shift={(-0.25,0)}] {} (dso);

\draw [Connection, bend left=53] (pros2) to  (dso);
\draw [Connection, bend left=42] (pros3) to (dso);

\draw [Connection, bend right=0] (solver) to node [midway, below left] {Ethereum} (contract);
\draw [Connection, bend right=15] (dso) to (contract);
\draw [Connection, bend right=0] (pros2) to (contract);
\draw [Connection, bend right=0] (pros3) to (contract);
\end{tikzpicture}%
\caption{Implementation view of the \Platform. A private Ethereum network (used for testing purposes) is the decentralized computation platform running the smart 
\ifExtended
\else
contracts.
\fi
}
\label{fig:components}
\end{figure}
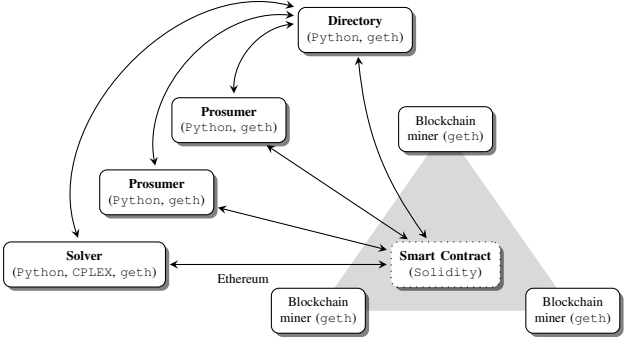

\begin{figure}[t]
\centering
\begin{tikzpicture}[x=1.5cm, y=0.9cm, font=\tiny,
  Component/.style={fill=white, draw, align=center, rounded corners=0.1cm, drop shadow={shadow xshift=0.05cm, shadow yshift=-0.05cm, fill=black}},
  Connection/.style={<->, >=stealth, shorten <=0.06cm, shorten >=0.06cm},
  Label/.style={midway, align=center, fill=white, fill opacity=0.75, text opacity=1}
]
  
\node [Component, dashed, minimum width=8cm] (ledger) at (0, 0.1) {\textbf{Distributed Ledger} (e.g., blockhain)};

\node [Component, align=center, minimum width=5.75cm] (sc) at (-0.75, 1) {\textbf{Smart Contract}\\(check offer and solution correctness, select best solution)};

\node [Component, dashed, minimum width=1.9cm, minimum height=0.55cm] (events) at (2, 1) {\textbf{Events}};

\node [Component, minimum width=1.8cm, minimum height=0.55cm] (solver) at (2, 3.75) {\textbf{Solver}};

\node [Component, dashed, minimum width=1.8cm, minimum height=0.55cm] (mixer) at (-2.1, 2) {\textbf{Anonymizing Mixer}\\(provide privacy)};

\node [Component, minimum width=1.8cm, minimum height=0.55cm] (prosumer) at (-2.1, 3.75) {\textbf{Prosumer}};

\node [Component, minimum width=1.8cm, minimum height=0.55cm] (directory) at (-0.5, 2.5) {\textbf{Directory}};

\draw [Connection] (prosumer) -- node [Label] {anonymous account} (mixer);

\draw [Connection, ->] (directory) -- node [Label] {close, finalize} (-0.5,1.35);

\draw [Connection, ->] (events) -- node [Label] {offers,\\closed} (solver);

\draw [Connection] (-1.6,3.4) -- node [Label, xshift=0.5cm] {connection addresses} (directory);

\draw [Connection, ->] (2,0.3) -- (2,0.7);
\draw [Connection, ->] (-2,0.3) -- (-2,0.7);
\draw [Connection, <-] (0,0.3) -- (0,0.7);

\draw [Connection, ->] (solver) -| node [Label] {potential\\solutions}  (1,1.35);

\draw [Connection, ->] (-1.5,3.6) -| node [Label] {offers}  (0.3,1.35);

\draw [Connection, <-] (-1.5,3.9) -- node [Label] {resource allocation for offers} (0.5,3.9) --  (1.5,1.35);
\end{tikzpicture}
    \caption{Data flow between actors of \Platform. 
        }
    \label{fig:tmp}
\end{figure}
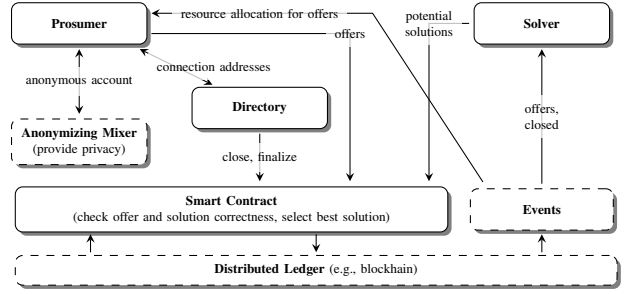

Figure \ref{fig:components} shows the key actors of our transaction management platform,
while
Figure \ref{fig:tmp} describes the data flow between these actors.
A directory actor provides a mechanism to look up connection endpoints, including the address of a deployed smart contract. In our previous work, we  described how to create a decentralized directory service using distributed hash tables~\cite{riaps1}. Therefore, we do not describe the implementation of this service any further in this paper. 
Prosumer actors (i.e., resource providers and consumers) post offers to the platform via functions provided by a smart contract. These functions check the correctness of each offer and then store it within the smart contract. An optional mixer service can be used to obfuscate the identity of the prosumers \cite{serial17}. By generating new anonymous addresses at random periodically, prosumers can prevent other entities from linking the anonymous addresses to their actual identities~\cite{Laszka17,serial17}, thereby keeping their activities private. 
Solver actors, which are pre-configured with constraints and an objective function, can listen to smart-contract events, which provide the solvers with information about offers. Solvers run at a pre-configured intervals, compute a resource allocation, and submit the solution allocation to the smart contract.
The directory, acting as a service director,  can then finalize a solution by invoking a smart-contract function, which chooses the best solution from all the allocations that have been submitted.
 Once a solution has been finalized, the prosumers are notified using smart-contract events.

\subsection{Scheduling}
\label{sec:scheduling}

Once the system is deployed,
providers and consumers will need to use it repeatedly for finding optimal resource allocations. 
For example, riders and drivers want to find optimal carpooling assignments every day, while users in an energy futures market want to find optimal energy trades every, e.g., 20 minutes.
Consequently, the platform has to gather offers and solve the resource allocation problem at regular time intervals.
Each one of these cycles is divided into two phases.
First, an \emph{offering phase}, in which providers and consumers can post new offers or cancel their existing offers (e.g., if they wish to change their offer based on changes in the market).
Second, a \emph{solving phase}, in which the resource allocation problem is solved for the posted (but not cancelled) offers.
At the end of the second phase, the assignments between providers and consumers are finalized based on the solution.
Then, a new cycle begins with an offering phase.

\subsection{Hybrid Solver Architecture}
\label{sec:solver}
The Resource Allocation Problem described in Section~\ref{sec:problem} can be solved by formulating it as an (integer) linear program (LP): feasibility constraints (Equations~\eqref{eq:feasible1} to~\eqref{eq:feasible4}) and constraint extensions (Section~\ref{sec:extConstr}) can all be formulated as linear inequalities, while the objective function (Equation~\eqref{eq:objective}) as well as the alternative objectives (Section~\ref{sec:extObj}) can be formulated as linear functions. Arguably, we could set up a solver actor that would receive offers from  prosumers, formulate a linear program, and use a state-of-the-art LP-solver (e.g., CPLEX~\cite{cplex2009v12}) to find an optimal solution. However, a simple N-to-1 architecture with N prosumers and 1 solver would suffer from the following problems: 
\begin{itemize}
    \item {\it Lack of trust in solver nodes:} Prosumers would need to trust that the solver is acting selflessly and is providing correct and optimal solutions. 
    \item {\it Vulnerability of the transaction management platform:} A single solver would be a single point of failure. If it were faulty or compromised, the entire platform would be faulty or compromised.  
    \item {\it Data storage:} For the sake of auditability, information about past offers and allocations should remain available even in case of node failures. 
\end{itemize}



A decentralized ledger with distributed information storage and consensus provided by blockchain solutions, such as Ethereum, is an obvious choice for ensuring the auditability of all events and providing distributed trust. However, computation is relatively expensive on blockchain-based distributed platforms\footnote{Further, Solidity, the preferred high-level language for Ethereum, currently lacks built-in support for certain features that would facilitate the implementation of an LP solver, such as floating-point arithmetics.}, solving the trading problem using a block\-chain-based smart contract would not be scalable in practice.
In light of this, we propose a \emph{hybrid implementation approach}, which combines the trustworthiness of blockchain-based smart contracts with the efficiency of more traditional computational platforms.

Thus, the key idea of our hybrid approach is to (1) use a high-performance computer to solve the computationally expensive linear program \emph{off-blockchain} and then (2) use a smart contract to record the solution \emph{on the blockchain}.
To implement this hybrid approach securely and reliably, we must address the following issues.
\begin{itemize}
\item Computation that is performed off-blockchain does not satisfy the auditability and security requirements that smart contracts do. Thus, the results of any off-blockchain computation must be verified 
by the smart contract before recording them on the blockchain.
\item Due to network disruptions and other errors (including deliberate denial-of-service attacks), the off-blockchain solver might fail to provide the smart contract with a solution on time (i.e., before assignments are supposed to be finalized). Thus, the smart contract must not be strongly coupled to the solver.
\item For the sake of reliability, the smart contract should accept solutions from multiple off-blockchain sources; however, these sources might provide different solutions.
Thus, the smart contract must be able to choose from multiple solutions (some of which may come from a compromised computer).
\end{itemize}



\subsection{Smart Contract}
\label{sec:smartcontract}

We implement a smart contract that can (1) verify whether a solution is feasible and (2) compute the value of the objective function for a feasible solution.
Compared to finding an optimal solution, these operations are computationally inexpensive, and they can easily be performed on a blockchain-based decentralized platform. Thus, we implement a smart contract that provides the following functionality:
\begin{itemize}
\item Solutions may be submitted to the contract at any time during the solving phase. 
The contract verifies the feasibility of each submitted solution, and if the solution is feasible (i.e., if it satisfies the constraint Equations \eqref{eq:feasible1} to \eqref{eq:feasible4}), then it computes the value of the objective function (i.e., Equation \eqref{eq:objective}).
The contract always keeps track of the best feasible solution submitted so far, which we call the \emph{candidate solution}.
\item At the end of the solving phase, the contract finalizes resource assignments for the cycle based on the candidate solution. If no solution has been submitted to the contract, then an empty allocation is used as a solution, which is always feasible but attains zero objective. 
\end{itemize}

This simple functionality achieves a high level of security and reliability.
Firstly, it is clear that an adversary cannot force the contract to finalize assignments based on an incorrect (i.e., infeasible) solution since such a solution would be rejected.
Similarly, an adversary cannot force the contract to choose an inferior solution instead of a superior one.
In sum, the only action available to the adversary is proposing a superior feasible solution, which would actually improve the transactive management platform.


To ensure that the smart-contract code is correct-by-construction~\cite{sifakis2013RSD}, we use the formal design environment FSolidM~\cite{mavridou2018designing} to design and generate the Solidity code of the smart contract. FSolidM allows designing Ethereum smart contracts as Labelled Transition Systems (LTS) with formal semantics. Each LTS can then be given to the NuSMV model checker~\cite{cimatti2002nusmv} to verify liveness, deadlock-freedom, and safety properties, which can capture important security concerns. 

\begin{figure}[t]
\centering
\includegraphics[width=0.8\columnwidth]{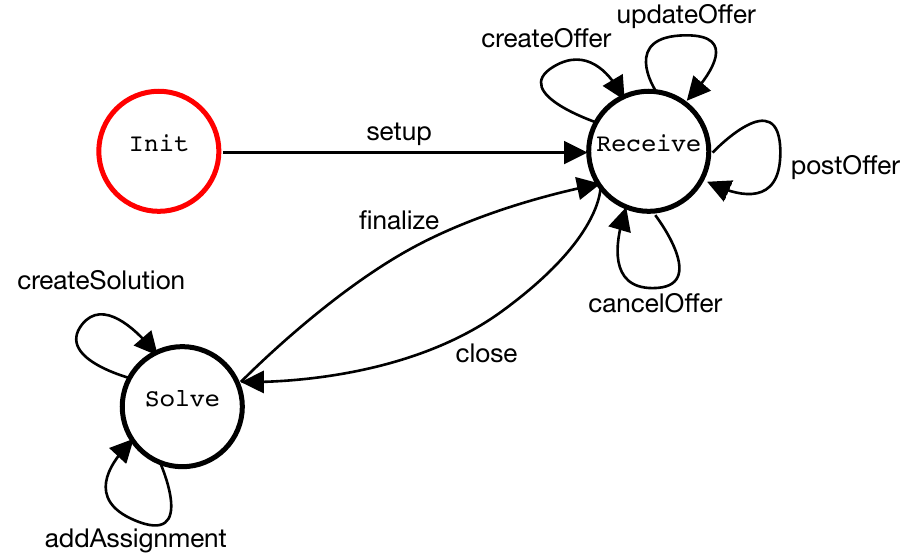}
\caption{FSolidM model of the \Platform smart contract.}
\label{fig:FSM}
\end{figure}


In Figure~\ref{fig:FSM}, we present the LTS representation of the transactive-platform smart contract, designed with FSolidM. The contract has three states:\footnote{Generated smart-contract code is not included in the paper because of space constraints. However, interested readers can view the code at \texttt{\url{https://github.com/visor-vu/transaction-management-platform}}}
\begin{itemize}[noitemsep, topsep=0pt, leftmargin=1em]
    \item \texttt{Init}, in which the contract has been deployed but  not been initialized. Before the contract can be used, it must be initialized (i.e., numerical parameters must be set~up).
    \item \texttt{Receive}, which corresponds to the \emph{offering} phase of a cycle (see Section~\ref{sec:scheduling}).
    In this state, prosumers may post (or cancel) their offers.
    \item \texttt{Solve}, which corresponds to the \emph{solving} phase of a cycle (see Section~\ref{sec:scheduling}). In this state, solvers may submit solutions (i.e., resource allocations) based on the posted (but not cancelled) offers.
\end{itemize}

In FSolidM, smart-contract functions are modeled as LTS transitions. Note that by design, each function may be executed only if the contract is in the origin state of the corresponding transition.
Our smart contract  has the following transitions (after the name of each transition, we list the function parameters):
\begin{itemize}[leftmargin=1em, noitemsep]
    \item from state \texttt{Init}:
    \begin{itemize}[noitemsep, leftmargin=0.5em]
        \item \texttt{setup(uint64 numTypes, uint64 precision, uint64 maxQuantity, uint64 lengthReceive, uint64 lengthSolve)}: initializes a contract with numerical parameter values, setting up the number of resource types, the arithmetic precision for calculations, the maximum quantity that may be offered, and the time length of the offering and solving phases; upon execution, the contract transitions to state \texttt{Receive}.
    \end{itemize}
    \item from state \texttt{Receive}:
    \begin{itemize}[noitemsep, leftmargin=0.5em]
        \item \texttt{createOffer(bool providing, uint64 misc))}: creates a blank offer (belonging to the prosumer invoking this transition) within the smart contract; parameter \texttt{providing} is true for providers and false for consumers, parameter \texttt{misc} is an arbitrary value that prosumers may use for their own purposes (e.g., to distinguish between their own offers); emits an \texttt{OfferCreated} event.
        \item \texttt{updateOffer(uint64 ID, uint64 resourceType, uint64 quantity, uint64 value)}: sets quantity and value for a resource type in an existing  offer (identified by the \texttt{ID} given in the \texttt{OfferCreated} event); may be invoked only by the entity that created the offer, and only if the offer exists but has not been posted yet; emits an \texttt{OfferUpdated} event.
        \item \texttt{postOffer(uint64 ID)}: posts an existing offer, enabling solvers to include this offer in a solution; may be invoked only by the entity that created the offer; emits an \texttt{OfferPosted} event.
        \item \texttt{cancelOffer(uint64 ID)}: cancels (i.e., ``unposts'') an offer, forbidding solvers from including this offer in a solution; may be invoked only by the entity that created the offer; emits an \texttt{OfferCanceled} event.
        \item \texttt{close()}: protected by a  guard condition on time, which prevents the execution of this transition before the offering phase of the current cycle ends; transitions to state \texttt{Solve}; emits a \texttt{Closed} event.
    \end{itemize}
    \item from state \texttt{Solve}:
    \begin{itemize}[noitemsep, leftmargin=0.5em]
        \item \texttt{createSolution(uint64 misc)}: creates a new, empty solution (i.e., resource allocation) within the smart contract; parameter \texttt{misc} is an arbitrary value that solvers may use for their own purposes (e.g., to distinguish between their own solutions); emits a \texttt{SolutionCreated} event.
        \item \texttt{addAssignment(uint64 ID, uint64 providingOfferID, uint64 consumingOfferID, uint64 resourceType, uint64 quantity, uint64 value)}: adds a resource assignment to an existing solution (identified by the \texttt{ID} given in the \texttt{SolutionCreated} event); may be invoked only by the entity that created the solution; checks a number of constraints ensuring that the solution remains valid if this assignment is added; emits an \texttt{AssignmentAdded} event.
        \item \texttt{finalize()}: selects the best solution and finalizes it by emitting an \texttt{AssignmentFinalized} event for each assignment in the solution; protected by a guard condition on time, which prevents the execution of this transition before the solving phase of the current cycle ends; transitions to state \texttt{Receive}.
    \end{itemize}
\end{itemize}

Notice that posting an offer or submitting a solution requires at least three or two function calls, respectively.
The reason for dividing these operations into multiple function calls is to ensure that the computational cost of each function call is constant:
\begin{itemize}[noitemsep, topsep=0pt]
    \item \texttt{createOffer}, \texttt{postOffer}, \texttt{cancelOffer}, and \texttt{createSolution} are obviously constant-cost.
    \item \texttt{updateOffer} adds a single resource type to an offer.
    \item \texttt{addAssignment} simply updates the sum amounts on the left-hand sides of Equations~\eqref{eq:feasible1} and~\eqref{eq:feasible2} for a single providing and a single consuming offer, respectively; and then it updates the sum in Equation~\eqref{eq:objective}.
\end{itemize}
%
With variable-cost functions, posting a complex offer or submitting a complex solution could be infeasible due to large computational costs, which could exceed the gas limit.\footnote{In Ethereum, each transaction is allowed to consume only a limited amount of gas, which corresponds to the computational and storage cost of executing the transaction.}

\begin{figure}[t]
    \centering
    \includegraphics[width=\columnwidth]{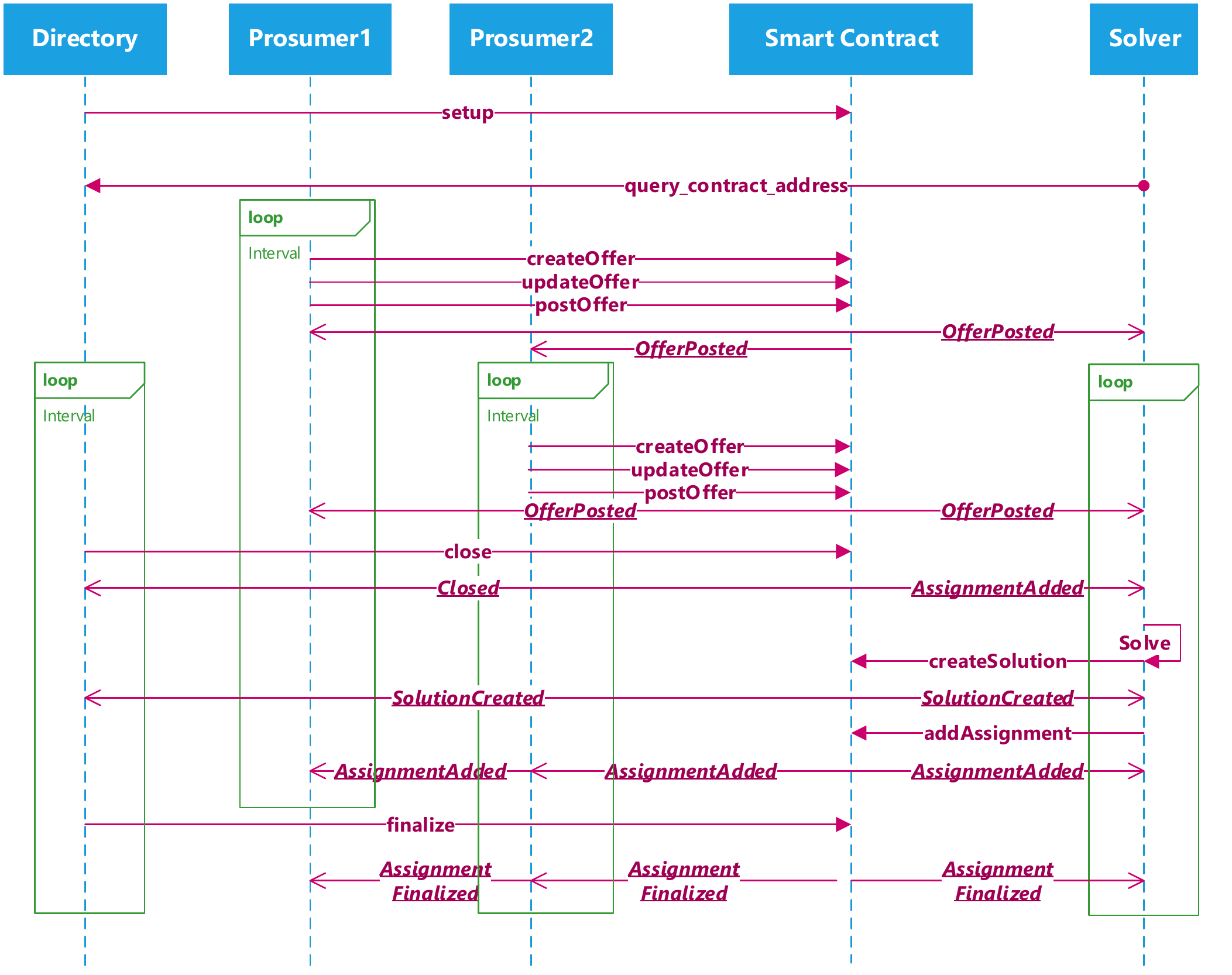}
    \caption{A possible sequence of operations in \Platform. Underlined text denotes events emitted by the smart contract. Some events, such as \texttt{OfferUpdated}, are omitted for simplicity. 
    }
    \vspace{-0.1in}
    \label{fig:workflow}
\end{figure}

 A typical sequence of function calls and events in \Platform is shown in Figure \ref{fig:workflow}.


\subsection{Analysis}
\label{sec:analysis}


\Aron{Note to self: extend!}
The computational cost of every smart-contract function is constant (i.e., $O(1)$) except for \texttt{finalize}, whose cost is an affine function of the size of the solution (i.e., $O(|A|)$).
Note that the cost of \texttt{finalize} depends on the size of the solution $A$ only because it emits an event for every assignment $\va \in A$.
These could be omitted for the sake of maximizing performance since the assignments have already been recorded in the blockchain anyway.
The number of function calls required for posting an offer depends on the number of resource types with non-zero quantity in the offer.
If there are $n$ such resource types, then $n + 2$ calls are required (\texttt{createOffer}, $n$ \texttt{updateOffer}, and \texttt{postOffer}).
The number of function calls required for submitting a solution $A$ is $1 + |A|$ (\texttt{createSolution} and $|A|$ \texttt{addAssignment}).

\subsubsection{Verification}
%
For the specification of safety and liveness properties, we use Computation Tree Logic (CTL)~\cite{baier2008principles}. CTL formulas specify properties of execution trees generated by transitions systems. The formulas are built from atomic predicates that represent 
transitions and statements of the
transition system (i.e., smart contract), using several operators, such as $\mathtt{AX}$, $\mathtt{AF}$, $\mathtt{AG}$ (unary) and,
$\mathtt{A}[\cdot\,\mathtt{U}\,\cdot]$,
$\mathtt{A}[\cdot\,\mathtt{W}\,\cdot]$ (binary).  Each operator
consists of a quantifier on the branches of the tree and a temporal modality, which together define when in the execution the operand sub-formulas must hold.  The intuition behind the letters is the
following: the branch quantifier is $\mathtt{A}$ (for ``All'') and the temporal modalities are
$\mathtt{X}$ (for ``neXt''), $\mathtt{F}$ (for ``some time in the
Future''), $\mathtt{G}$ (for ``Globally''), 
$\mathtt{U}$ (for ``Until'') and $\mathtt{W}$ (for ``Weak until'').  
A property is satisfied if it holds in the initial state of the transition systems.  
For instance, the formula
$\mathtt{A}[p\,\mathtt{W}\,q]$ specifies that in \emph{all execution
  branches} the predicate $p$ must hold \emph{up to the first state}
(not including this latter) where the predicate $q$ holds.  
Since we used the weak until operator $\mathtt{W}$, if $q$ never
holds, $p$ must hold forever.
As soon
as $q$ holds in one state of an execution branch, $p$ does not need to hold
anymore, even if $q$ does not hold.  On the contrary, the formula
$\mathtt{AG}\,\mathtt{A}[p\,\mathtt{W}\,q]$ specifies that the
subformula $\mathtt{A}[p\,\mathtt{W}\,q]$ must hold in \emph{all
  branches at all times}.  Thus, $p$ must hold whenever $q$ does not hold, i.e., $\mathtt{AG}\,\mathtt{A}[p\,\mathtt{W}\,q] = \mathtt{AG}\,(p \lor q)$.

We verified correctness of behavioral semantics with the NuSMV model checker~\cite{cimatti2002nusmv}, by verifying the following properties:
\begin{itemize}[noitemsep, leftmargin=0.9em]
    \item \emph{deadlock-freedom}, which ensures that the contract cannot enter a state in which progress is impossible;
    \item \emph{``if \texttt{close} happens, then \texttt{postOffer} or \texttt{cancelOffer} can happen only after \texttt{finalize}''}, translated to CTL as: $\mathtt{AG}(\texttt{close})$ $\rightarrow$ $\mathtt{AX}$ $\mathtt{A}$ $[\neg ($\texttt{postOffer} $\wedge$ \texttt{cancelOffer}) $\mathtt{W}$ \texttt{finalize}$]$, which ensures that prosumers cannot post or cancel their offers once the solvers have started working;
    \item \emph{``\texttt{OfferPosted(ID)} can happen only if \texttt{(ID < offers.length) \&\& !offers[ID].posted \&\& (offers[ID].owner == msg.sender)}''}, translated to CTL as:\\ $\mathtt{A}[ \neg$\texttt{OfferPosted(ID)} $\mathtt{W}$  \texttt{(ID < offers.length) \&\& !offers[ID].posted \&\& (offers[ID].owner == msg.sender)}$]$, which ensures that an offer can be posted only if it has been created (but not yet posted) and only by its creator;
    \item \emph{``\texttt{OfferCanceled(ID)} can happen only if  \texttt{(ID < offers.length) \&\& offers[ID].posted \&\& (offers[ID].owner == msg.sender)}''}, translated to CTL as:\\ $\mathtt{A}[ \neg$\texttt{OfferCanceled(ID)} $\mathtt{W}$ \texttt{(ID < offers.length) \&\& offers[ID].posted \&\& (offers[ID].owner == msg.sender)}$]$, which ensures that an offer can be canceled only if it has been posted and only by the poster;
    \item \emph{``if \texttt{finalize} happens, then \texttt{createSolution} can happen only after \texttt{close}''}, translated to CTL as:\\ $\mathtt{AG}(\texttt{finalize})$ $\rightarrow$ $\mathtt{AX}$ $\mathtt{A}$ $[\neg$\texttt{createSolution} $\mathtt{W}$ \texttt{close}$]$, which ensures that solutions can be posted only in the solving phase.
\end{itemize}




\section{Case Studies}
\label{sec:results}



To evaluate our platform, we present two case studies, based on the energy trading and carpooling problems (Section~\ref{sec:ExaApp}), with numerical results. 
The computational results for the carpool example were obtained on a virtual machine configured with 16 GB of RAM and 4 cores of a i7-6700HQ processor. The energy market example results were obtained on a virtual machine configured with 8GB of RAM and 2 cores of an i7-6700HQ processor. For these experiments, we used
 our private Ethereum blockchain network~\cite{EthereumBook}. 

\subsection{Carpooling Problem}
\label{sec:carpool}

In this section, we describe a simulated carpooling scenario. The problem of carpooling assignment was introduced earlier in Section~\ref{sec:carpoolingprob}.
Here, we model a carpool prosumer as an actor that specifies
\begin{enumerate*}
    \item whether it is providing or requesting a ride,
    \item the number of seats being offered/requested,
    \item a residence,
    \item a destination,
    \item a time interval during which the ride is available/required,
    \item and a radius specifying how far out of their way they are willing to travel.
\end{enumerate*}
To setup the carpooling problem, we need to identify these parameters and encode them as offers. 

\begin{figure}[t]
\centering
\includegraphics[width=0.7\columnwidth]{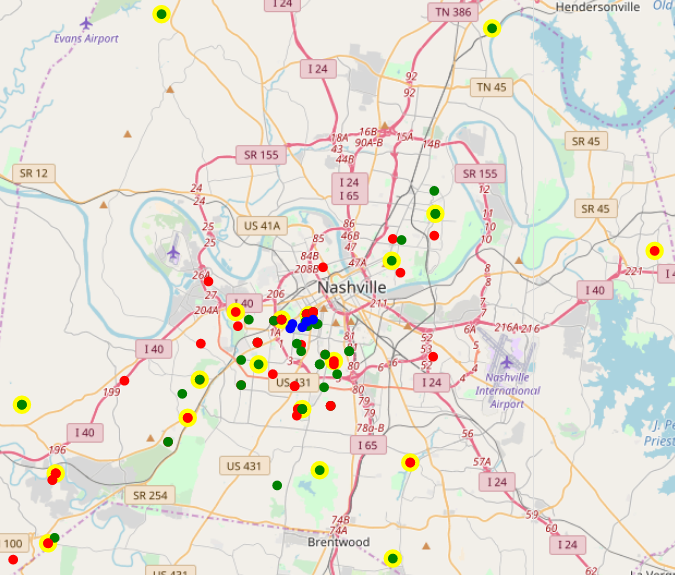}
\caption{
Green and red dots mark the 75 residences (anonymized and resampled). Blue dots are destinations on campus. We used K-Means to identify 20 central locations (yellow dots) for pickup.
}
\label{fig:clusters}
\end{figure}

Residences were generated by sampling from real-trip distribution data of Vanderbilt 
\ifExtended
\else
University.
\fi
Destinations were chosen uniformly at random for each prosumer from the~5 garages around Vanderbilt University.
Other parameters were also chosen randomly:  number of seats from the range of 1 to 3,  prosumer type from producer or consumer,  time interval from 15-minute intervals between 7:00 and 9:30AM. The ``out of the way'' metric was chosen to be half of the distance between the residence and the destination. For a provider, the center of the pick-up circle is the midway point between the residence and the destination, and for a consumer,  the center is the residence.

Since each prosumer has a distinct residence, encoding it as a unique resource type would mean that every  prosumer would need to have the address of every other prosumer to determine if they are in their pick-up range. Instead, we specify \textit{pick-up points} which are public locations were carpoolers can meet. Each prosumer can determine which pick-up points are within their out-of-the-way radius and list those points in their offer. To encode these values, we assign an ID to each pickup point and destination. Finally, we encode each 15-minute interval using a timestamp. 

An offer consists of a collection of alternative resource types, each with a quantity and value.
We encode a resource type, which is a combination of a time interval, a pick-up point, and  a destination, as a 64-bit unsigned integer. 
For example, if timestamp is 1523621700,  pick-up location ID is 15, and destination ID is 3, then the resource type is 1523621700153. 
A complete offer may look as follows: 
\begin{align*}
      \{ \texttt{True}, ~ 
     & \{1523623500173: 2, 
        1523623500153: 2, \\
     &  ~\,1523624400153: 2, 
        1523624400173: 2\}, \\
     & \{1523623500173: 10, 
        1523623500153: 10, \\
     & ~\,1523624400153: 10, 
        1523624400173: 10\} ~\} .
\end{align*}
%
In this offer, the prosumer is offering rides (\texttt{True} for providing), has two pick-up locations in range (17 and 15), drives to destination 3, is available in two time intervals, offers 2 seats, and asks for value 10 in exchange for a ride. 

In our experiment, we selected 75 prosumers for the carpool service simulation. The red and green points in Figure~\ref{fig:clusters} are the locations of the consumers and producers randomly sampled from the anonymized distribution data of employees of Vanderbilt University. The yellow points were selected as pick up locations using K-Means clustering choosing 20 clusters. The blue points are 5 garages around Vanderbilt campus where employees typically park.  

\begin{figure}[t]
\includegraphics[width=\columnwidth]{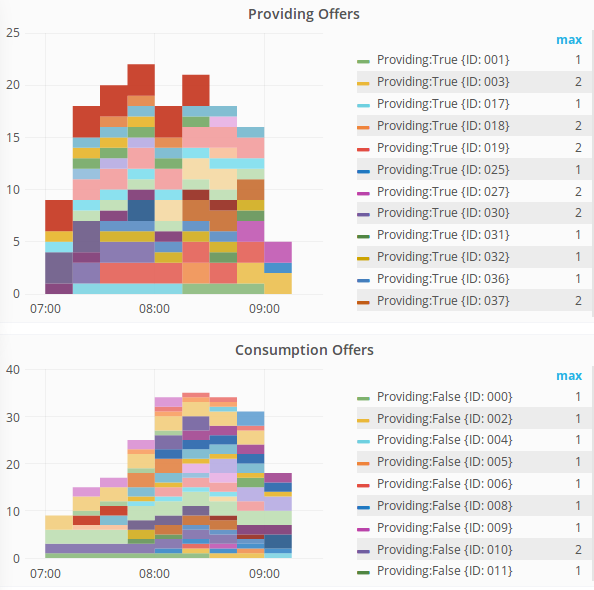}
\caption{Carpooling offers posted.}
\label{fig:offers}
\end{figure}

Figure \ref{fig:offers} shows all the offers posted to the carpool platform. Each color is a unique offer. For example, the providing offer that is represented by red bars on the first six columns (7:00--8:15AM) having a height of 2, offers 2 seats with pick-up any time between 7:00AM and 8:15AM. The offers are stacked, showing how many seats are potentially available at that time. The chart combines all start and end locations; however, these could be separated. 

\begin{figure}[t]
\centering
\includegraphics[width=0.9\columnwidth]{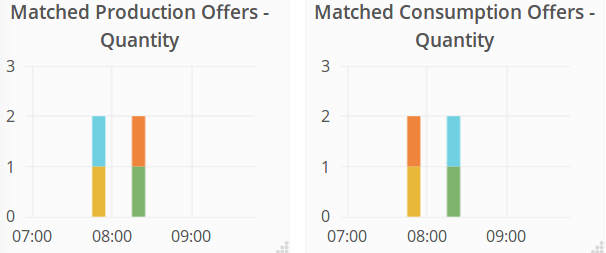}
\caption{Matched (i.e., assigned) offers in the solution.}
\label{fig:matches}
\end{figure}

Figure \ref{fig:matches} shows the offers matched in each interval. For example, at 7:45AM, 2 providing offers--yellow (1 seat) and blue (1 seat)--and two consuming offers--yellow (1 seat) and orange (1 seat)--were matched. These are again grouped only by time in the figure, and not by start or end points. 
The running time of the solver was $23$ ms, while the time between the request for finalization and emission of \texttt{AssignmentFinalized} events was $29$ s.
%



\begin{figure}[t]
    \centering
    \includegraphics[width=0.95\linewidth]{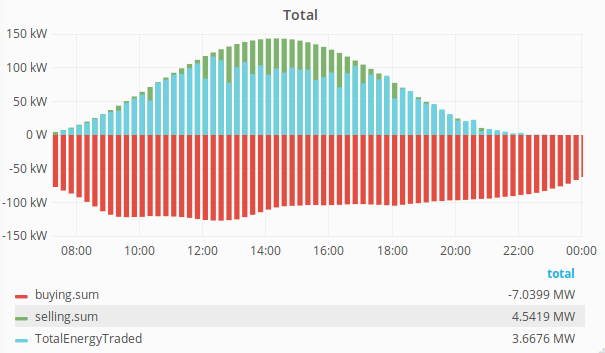}
    \caption{Total energy production capacity (green) and energy demand (red) for each interval, as well as the total energy traded in each interval (blue).}
    \label{fig:totals}
\end{figure}

\begin{figure} [h]
    \centering
    \includegraphics[width=0.9\linewidth]{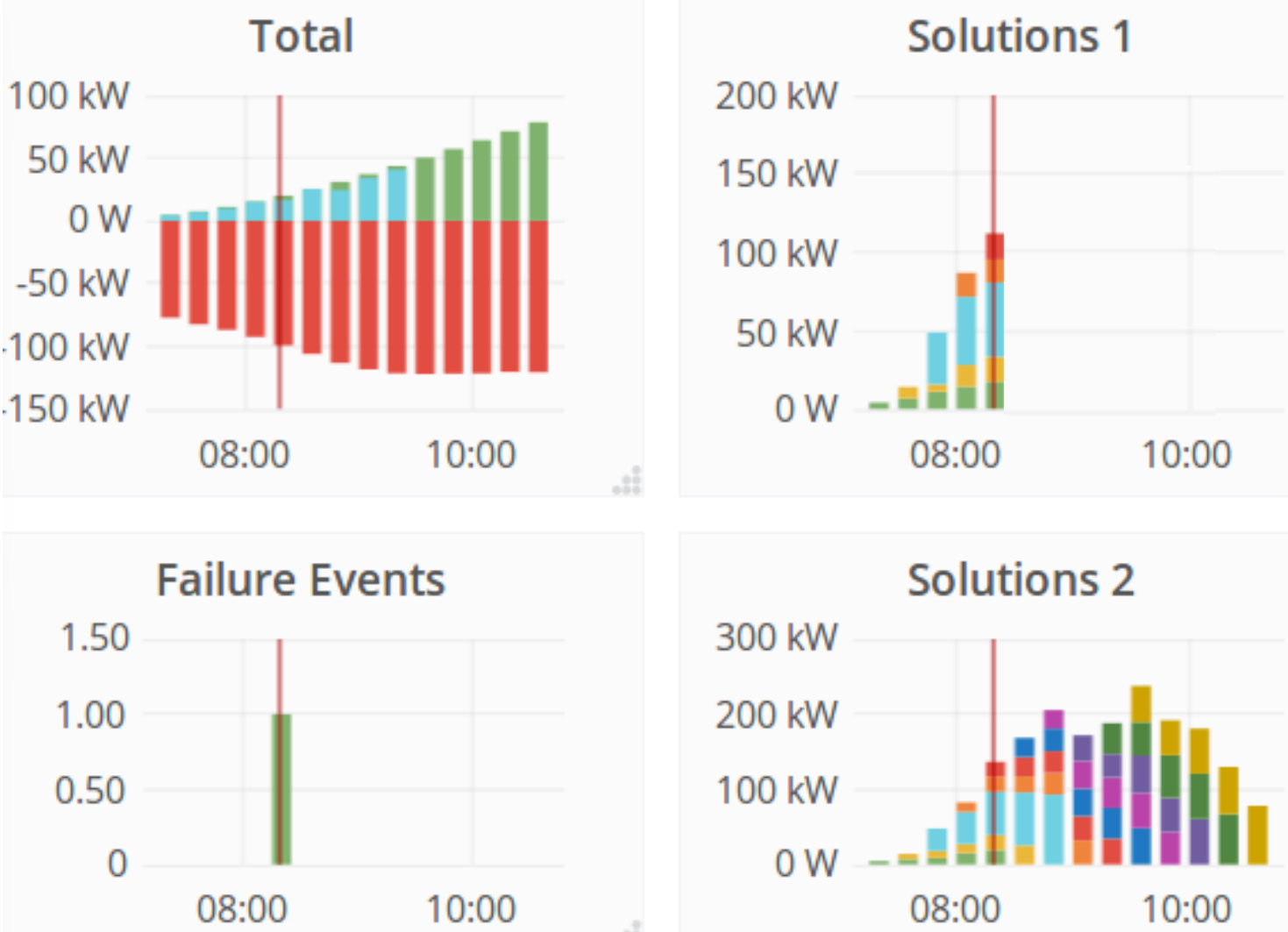}
    \caption{A failure scenario with failure at 8:15AM. The solver can submit new solutions as time progresses; the most recent solution is the color that is on the top of the stack for an interval.}
    \label{fig:failure}
    \vspace{-0.05in}
\end{figure}


\subsection{Energy Trading Problem}
 \label{sec:energy}

  To show the versatility of our transaction management platform, we now apply it to the problem of energy trading  within a microgrid, which we introduced in Section~\ref{sec:energyFuturesMarket}.
  In this example, a prosumer is modeled as an actor with an energy generation and consumption profile for the near future. In practice, the generation profile would be typically derived from predictions based on the weather, energy generation capabilities, and the amount of battery storage available. The consumption profile would be derived from flexible power loads, like  washers and electric vehicles.  

To represent future generation or consumption at a certain time, 
resource types encode 
timestamps for 15-minute intervals, during which the power will be generated or consumed. 
As an example, consider a battery that has 500 Wh energy, which could be discharged any time between 9AM and 10AM.
This can be represented by an offer having resource types 900, 915, 930, and 945, specifying a quantity of 500 Wh for each.

 For our simulation, the prosumer energy profiles are load traces recorded by Siemens during a day from a microgrid in Germany, containing $102$ homes 
 ($5$ producers and $97$ consumers). Since the dataset does not include prices, we assume reservation prices to be uniform in our experiments, and focus on studying the amount of energy traded and the performance of the system. 


Figure~\ref{fig:totals} shows the total production (green) and consumption (red) across this microgrid, as well as the total energy traded per interval (blue) using our platform. The horizontal axis shows the starting time for each of the $96$ intervals.



In another simulation, we exercise the hybrid solver architecture by running multiple solvers, and after some time, cause one to fail. This result is shown in Figure~\ref{fig:failure}. The narrow vertical red line indicates when solver 1 fails at 8:15AM. Up until that point, solver 1 submitted the green, yellow, light blue, orange solutions, with the final solution being red.
On the other hand we see that solver 2 continues to provide solutions for later intervals. 





 
\section{Related Research}
\label{sec:related}

\textbf{Online Information Management Platforms:}
Smart and connected community systems are designed to collect, process, transmit, and analyze data. In this context, data collection usually happens at the edge because that is where edge devices with sensors are deployed to monitor surrounding environments. \Platform does not suggest a specific data collection methodology. Rather, it follows an actor-driven design pattern where ``prosumer'' actors can integrate their own agents into \Platform by using the provided APIs.  Another concern of these platforms is the cost of processing. Traditionally, this problem was solved using scalable cloud resources in-house~\cite{schmidt2014elastic}. However, \Platform enables a decentralized ecosystem, where components of the platform can run directly on edge nodes, which is one of the reasons why we designed it to be asynchronous in nature.

To an extent, the information architecture of \Platform can be compared to  dataflow engines~\cite{Storm, Spark, neumeyer2010s4}. All of these existing dataflow engines use some form of a computation graph, comprising computation nodes and dataflow edges. These engines are designed for batch-processing and/or stream-processing high volumes of data in resource intensive nodes, and do not necessarily provide additional ``platform services'' like trust management or solver nodes. 

\textbf{Integration with Blockchains:}
\Platform integrates a blockchain because it enables the digital representation of resources, such as energy and financial assets, and their secure transfer from one  party to another. Further, blockchains constitute an immutable, complete, and fully auditable record of all transactions that have ever occurred in the  system. This is in line with the increased interest and commercial adoption of blockchains~\cite{TheTruthAdoption:online}, which 
has yielded market capitalization surpassing \$75 billion USD~\cite{BitcoinPrices:online} for Bitcoin and \$36 billion USD for Ethereum ~\cite{EthereumPrices:online}. Prior work has also considered the security and privacy of IoT and Blockchain integrations~\cite{dorri2017blockchain,ouaddah2017towards,christidis2016blockchains}. 

The biggest challenge  in these integrated systems comes from computational-complexity limitations and from the complexity of the consensus algorithms. 
In particular, their transaction-confirmation time is relatively long and variable, primarily due to the widely-used proof-of-work algorithm. 
Further, blockchain-based  computation is relatively expensive, which is the main reason why we  separated finding a solution and validating the solution into two separate components in \Platform.

\textbf{Correctness of Smart Contracts:}
Both verification and automated vulnerability discovery are considered in the literature for identifying smart-contract vulnerabilities. 
For example, Hirai performs a formal verification of a smart contract that is used by the Ethereum Name Service~\cite{hirai2016formal}.
However, this verification proves only one particular property and it involves relatively large amount of manual analysis.
In later work, Hirai defines the complete instruction set of the Ethereum Virtual Machine in Lem, a language that can be compiled for interactive theorem provers~\cite{hirai2017defining}.
Using this definition, certain safety properties can be proven for existing contracts.

Bhargavan et al.\ outline a framework for verifying the safety and correctness of Ethereum smart contracts~\cite{bhargavan2016short}.
The framework is built on tools for translating Solidity and Ethereum Virtual Machine bytecode contracts into $F^*$, a functional programming language aimed
at program verification.
Using the~$F^*$ representations, the framework can verify the correctness of the Solidity-to-bytecode compilation as well as detect certain vulnerable patterns. Luu et al.\ provide a tool called \textsc{Oyente}, which can analyze smart contracts and detect certain typical security vulnerabilities~\cite{luu2016making}. The main difference between prior work and the approach that we are using (i.e., verifying FSolidM models with NuSMV) is that the former can prevent a set of typical vulnerabilities, but they are not effective against vulnerabilities that are atypical or belong to types which have not been identified yet.

\section{Conclusion}
\label{sec:conclusion}



Smart and connected community applications require decentralized and scalable platforms due to the large number of participants and the lack of mutual trust between them. In this paper,  we introduced a transactive platform for resource allocation, called \Platform. We first formulated a general problem that can be used to represent a variety of resource allocation problems in smart and connected communities. Then, we described an efficient and trustworthy platform based on a hybrid approach, which combines the efficiency of traditional computing environments with the trustworthiness of blockchain-based smart contracts. Finally, we demonstrated the applicability of our platform using two case studies based on real-world data.


\vspace{0.30em}
\noindent\textbf{Acknowledgement:}
This work was funded in part by a grant from Siemens, CT and in part by a grant from NSF under award number CNS-1647015.
\vspace{-0.1em}

\let\oldbibliography\thebibliography
\renewcommand{\thebibliography}[1]{\oldbibliography{#1}
\setlength{\itemsep}{0pt}} 

\bibliographystyle{IEEEtran}
\small{
\bibliography{references} 
}


\end{document}